\documentclass[conference,10pt,phv]{IEEEtran}
\usepackage{cite}
\usepackage{amsmath,amssymb,amsfonts}
\usepackage{algorithmic}
\usepackage{graphicx}
\usepackage{siunitx}
\graphicspath{{artwork/}} 
\usepackage{textcomp}
\usepackage{xcolor}
\usepackage{fancyhdr}

\def\BibTeX{{\rm B\kern-.05em{\sc i\kern-.025em b}\kern-.08em
    T\kern-.1667em\lower.7ex\hbox{E}\kern-.125emX}}

\usepackage{booktabs}
\usepackage[pdfauthor={Tim Prangemeier},hidelinks, bookmarks={false}, draft]{hyperref}

\usepackage{soul}
\usepackage[outline]{contour}

\definecolor{tab:TUred9a}{RGB}{233, 80, 62}
\definecolor{tab:TUgreen3a}{RGB}{80, 182, 149}
\definecolor{tab:TUblue1a}{RGB}{93, 133, 195}
\definecolor{tab:tud0c}{cmyk/RGB/HTML}{0,0,0,.6/137,137,137/898989}

\definecolor{tab:darkviolet}{RGB}{148, 0, 211}
\definecolor{tab:bggrey}{RGB}{163, 163, 163}
\definecolor{tab:trapblack}{RGB}{0, 0, 0}
\definecolor{tab:hotpink}{RGB}{255, 105, 180}
\definecolor{tab:plum}{RGB}{221, 160, 221}
\definecolor{tab:violet}{RGB}{238, 130, 238}

\newcommand \colorindicator[2]{%
	\begingroup%
	\setul{0.25ex}{0.4ex}%
	\contourlength{0.2ex}%
	\setulcolor{#1}%
	{{\phantom{#2}}}\llap{\contour{white}{#2}} \textcolor{#1}{\tiny{$\blacksquare \hspace{-.5mm} \blacksquare$}}%
	\endgroup%
}

\usepackage[colorinlistoftodos]{todonotes}

\newcommand{\tds}[3][]{
	\todo[color=gray!40,inline,#1,caption={#2},disable] {\underline{#2:} \\#3} }
\newcommand{\tdsDoneHide}[3][]{\todo[color=green!10,inline,#1,caption={$\checkmark$ #2},disable]{$\checkmark$ #2} }

\newcommand{\tdsCamera}[3][]{
	\todo[color=gray!40,inline,#1,caption={Camera Ready: #2},nolist,disable] {\underline{Camera Ready To Do:}\\ #2: \\#3} }

\begin{document}
\pagenumbering{roman}

\title{Multiclass Yeast Segmentation in Microstructured Environments with Deep Learning}

\author{ 
\IEEEauthorblockN{
	Tim Prangemeier, 
	Christian Wildner, 
	Andr\'e O. Fran\c{c}ani, 
	Christoph Reich, 
	Heinz Koeppl\IEEEauthorrefmark{5}}
\vspace{0.05in}\IEEEauthorblockA{Centre for Synthetic Biology,\\Department of Electrical Engineering and Information Technology,\\
	Department of Biology,\\
	Technische Universit\"at Darmstadt\\
	\emph{\IEEEauthorrefmark{5}heinz.koeppl@bcs.tu-darmstadt.de}}
}

\maketitle
\thispagestyle{plain}
\fancypagestyle{plain}{
	\fancyhf{} 
	\fancyfoot[L]{978-1-7281-9468-4/20/\$31.00~\copyright2020~IEEE, CIBCB 2020} 
	\renewcommand{\headrulewidth}{0pt}
	\renewcommand{\footrulewidth}{0pt}
}
\begin{abstract}
Cell segmentation is a major bottleneck in extracting quantitative single-cell information from microscopy data. The challenge is exasperated in the setting of microstructured environments. While deep learning approaches have proven useful for general cell segmentation tasks, existing segmentation tools for the yeast-microstructure setting rely on traditional machine learning approaches. Here we present convolutional neural networks trained for multiclass segmenting of individual yeast cells and discerning these from cell-similar microstructures. We give an overview of the datasets recorded for training, validating and testing the networks, as well as a typical use-case. We showcase the method's contribution to segmenting yeast in microstructured environments with a typical synthetic biology application in mind. The models achieve robust segmentation results, outperforming the previous state-of-the-art in both accuracy and speed. The combination of fast and accurate segmentation is not only beneficial for \emph{a posteriori} data processing, it also makes online monitoring of thousands of trapped cells or closed-loop optimal experimental design feasible from an image processing perspective.
\end{abstract}



\section{Introduction}
\label{sec:introduction}
\addcontentsline{tdo}{todo}{\textbf{Introduction}}
\setcounter{page}{1}
\pagenumbering{arabic}

\tdsDoneHide{spelling U-Net (Falk); Ronneberger less strict U/u-net}{}

\tdsDoneHide{SYNBIO ETC IS IMPORTANT! KEEP IN< LESSEN FOCUS}{}
\tdsDoneHide{our applciaiton is synbio characterisation, but method can be used beyond that?}{}
Time-lapse fluorescence microscopy (TLFM) is a powerful technique for studying cellular processes in intact living cells. For example, it has lead to insights into gene regulation and into the role of cell heterogeneity in cancer therapeutics \cite{Rosenfeld2005,Burke2016,VanValen2016}. The vast amount of quantitative and standardised data TLFM yields, promises to constitute the backbone for rational design of \emph{de novo} biomolecular functionality with quantitatively predictive modelling in synthetic biology \cite{Pepperkok2006,Xiang2018}. Ideally, well characterised parts and modules are combined \emph{in silico} in a bottom up approach \cite{Xiang2018,Bittihn2018,Gomez2019,Prangemeier2020}, for example to detect and kill cancer cells \cite{Xie2011,Si2018}. Concurrently accounting for cell-to-cell variability and biomolecular circuit dynamics on the single-cell level are key advantages of TLFM \cite{Leygeber2019,Kraus2017,Lugagne2020}, enabling the mathematical reconstruction of intracellular processes \cite{Zechner2014,Bakker2017,Prangemeier2018}.  

\begin{figure}[htbp]
	\centerline
	{\includegraphics{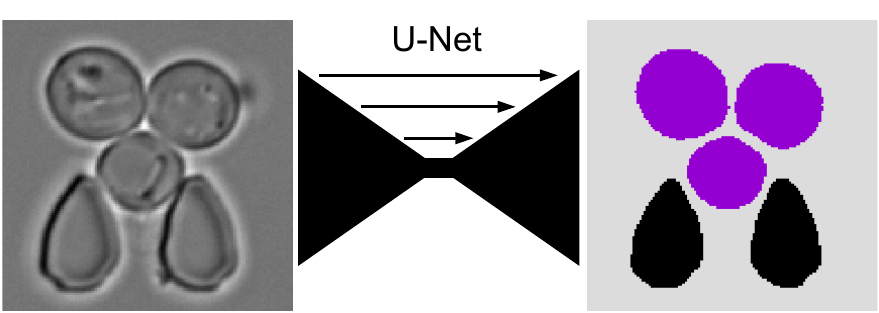}}
	\caption[$\blacksquare$ Fig. \ref{fig:intro_schem} : simple intro schematic (\checkmark)]{Schematic of U-Net semantically segmenting microscope imagery into multiple classes and discerning yeast cells (violet) from  trap microstructures.}
	\label{fig:intro_schem}
\end{figure}

A typical TLFM experiment with high-throughput microfluidics yields thousands of specimen images requiring automated segmentation, examples include \cite{Schneider2017,Hofmann2019,Bakker2017,Crane2014}. This is schematically depicted in Fig. \ref{fig:intro_schem}, where three yeast cells are held in place by microscopic trap structures. Variants of this approach exist for both of the model organisms of choice in synthetic biology, \emph{Escherichia coli} and the yeast \emph{Saccharomyces cerevisiae}, which we consider here. Segmenting each individual cell enables pertinent information about its properties to be extracted quantitatively. For example the abundance of a fluorescent reporter molecule can be measured, giving insight into the inner workings of the cell.
\tdsDoneHide{where to add model organisms yeast and Ecoli?}{}
\tdsDoneHide{The model organisms of choice in synthetic biology are yeast (\emph{Saccharomyces cerevisiae}) and \emph{Escherichia coli}}{}
\tdsDoneHide{they are correct: citation brackets are poorly formatted}{}
%
%
%
%
Segmentation is a major bottleneck in quantifying single-cell microscopy data and manual analysis is prohibitively labour intensive  \cite{VanValen2016,Bakker2017,Versari2017,Falk2019,Sauls2019,Lugagne2020,Gupta2019}. Curation time and accuracy is not only a drawback on the amount of experiments that can be performed, but also limits the type of experiments \cite{VanValen2016,Moen2019}. For example, harnessing the potential of advanced closed-loop optimal experimental design techniques \cite{Gomez2019,Prangemeier2018} requires online monitoring with fast segmentation capabilities. An additional challenge in the yeast-trap configuration depicted in Fig. \ref{fig:intro_schem} is that cells and microstructures need to be discerned, for which the vast majority of segmentation methods are not designed. Deep learning methods such as the U-Net \cite{Ronneberger2015,Falk2019} are increasingly outperforming conventional machine learning, however, they have not yet been specifically adapted for the segmentation scenario presented here (Fig. \ref{fig:intro_schem}).
\tdsDoneHide{add discerning traps and cells here more!}{}


\tdsCamera{refine images? (pixelation due to $128\times128$)}{some images are clearer than others, this is out of python already, can only be combatted by using labels instead of predictions and the original resolution (before going down to 128x128)}
\tdsDoneHide{yes: decide: make both seg schematic and complex?}{}
\tdsCamera{Note:the efficacy of yeast image cytometry hinges on the accurate/ robust high-throughput segmentaton, while much of the future potential pivots on the online monitoring capabilities, in particular interactive experimentation requiring real-time segmentaiton capabilities (maybe taking somewhat lesser accuracy into account)}{}
\tdsDoneHide{add lit: papers for heterogeneity, dynamics and traps}{}
\tdsDoneHide{add lit: importance of TLFM sources from COBIOT}{}

%

\tdsDoneHide{add: gap is U-Net on trapped yeast}{}
\begin{figure*}[htbp]
	\centerline{\includegraphics[width=1\textwidth]{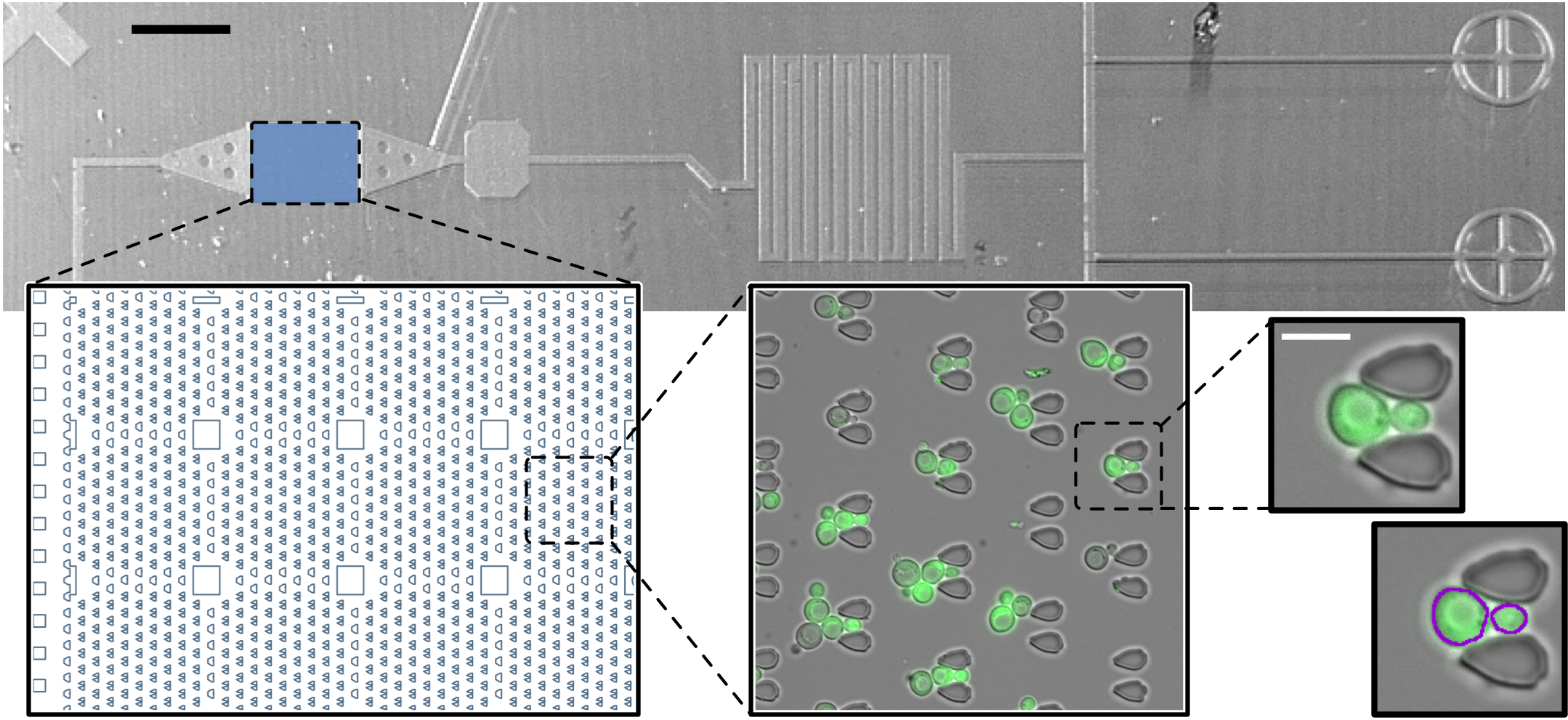}}
	\caption[$\blacksquare$ Fig. \ref{fig:complex}: complex whole chip \checkmark]{Microscope images and trap chamber design of the microstructured environment for hydrodynamically trapping yeast for concurrent long-term culture and imaging. The trap chamber (blue) contains an array of approximately \SI{1000}{} traps. A microscope records numerous positions automatically. Single specimen images show a pair of microstructures and  fluorescent cells (green), violet contours indicate segmentation; black scale bar \SI{1}{\milli\meter}, white scale bar \SI{10}{\micro\meter}.}
	\label{fig:complex}
\end{figure*}
\tdsCamera{finish complex, add scale bars and finish caption}{}
\tdsDoneHide{add: gap is U-Net on trapped yeast}{}
In this study, we address the automated segmentation bottleneck for the configuration of yeast cells in microstructured environments. We present a modified U-Net architecture for multiclass segmentation of both yeast cells and traps. Section \ref{sec:background} introduces the microstructured environment and the recent state-of-the-art segmentation methods for this setting. We present the annotated dataset, the experimental setup for data acquisition, as well as network architectures and training in Section \ref{sec:method}. In Section \ref{sec:results}, we analyse network performance and compare this with previously reported methods. We interpret the results in Section \ref{sec:discussion} and highlight their potential for processing time-lapse fluorescent microscopy data. Our models exceed the current state-of-the-art for this cell-microstructure configuration significantly in accuracy (measured by the intersection-over-union) and speed. We effectively maintain segmentation performance, while reducing the number of parameters to $1/6$-th of the original. Segmentation runtimes are significantly reduced, enabling both higher throughput \textit{a posteriori} data processing and making online monitoring of experiments with approximately $1000$ cell traps feasible. We summarise our study and highlight key conclusions in Section \ref{sec:conclusion}.  
\tdsDoneHide{shorten?}{}
\tdsDoneHide{tone down?!}{}
\tdsDoneHide{add that on par with humans and we alleviate the need to manual curation?}

\tdsDoneHide{add CIBCB refrences?}{}

\tdsDoneHide{add italics first mention of yeast: \emph{Saccharomyces cerevisiae}}{}

\tdsCamera{Note: the efficacy of yeast image cytometry hinges on the accurate/ robust high-throughput segmentaton, while much of the future potential pivots on the time-lapse capabilities, in particular interactive experimentation requiring real-time segmentaiton capabilities (maybe taking somewhat lesser accuracy into account)}{}



\section{Background}
\label{sec:background}
\addcontentsline{tdo}{todo}{\textbf{Background}}
\tdsDoneHide{spelling U-Net (Falk); Ronneberger less strict U/u-net}{}


\subsection{Microstructured yeast culture environments}

The cell-microstructure configuration we consider here, is designed for long-term culture of yeast cells in tightly controlled conditions within the focal plane of a microscope \cite{Crane2014}. Examples of its routine employ include Fig. \ref{fig:complex} and \cite{Prangemeier2020,Hofmann2019,Bakker2017,Schneider2017,Leygeber2019}. Cells are hydrodynamically trapped by a constant flow of media, ensuring their long-term growth and allowing the introduction of chemical perturbations. The microfluidic chips typically comprise \SI{1000}{} trap microstructures per trap chamber. An automated microscope records an entire trap chamber by imaging both the brightfield and fluorescent channels at approximately \SI{20}{} neighbouring positions. Time-lapse recordings of this configuration allow individual cells to be tracked through time, with robust segmentation facilitating tracking \cite{Moen2019,Gupta2019,Lugagne2020}. Tracking can itself be a limiting factor with regard to the data yield of an experiment \cite{Versari2017}.
\tdsCamera{split tracking part off?}{}
\tdsCamera{Note: Microfluidics now makes it possible to cultivate cells in constant conditions for a long period of time and image cytometry allows to delve into the dynamics of their inner workings}{}
\tdsDoneHide{add Crane2014 reference}{}


\tdsDoneHide{figure intro HK from meeting: make complex, whole chip }{
	$\bullet$ make it complex, best would be to use whole chip overview, then imaging chamber, then position then trap\\
	$\bullet$ add fluorescence image, fluorescence readout?}

%

\subsection{Single-cell segmentation tools}
\label{ssec:BG_segmentation}
\addcontentsline{tdo}{todo}{\textbf{\nameref{sec:introduction}: \nameref{ssec:BGsegmentation}}}

\label{sssec:yeast_seg}
An extensive body of research into the automated processing of microscopy imagery dates back to the middle of the 20-th century \cite{Gupta2019}. Recently, deep learning methods are increasingly outperforming conventional machine learning; comprehensive reviews of their application to biomedical or TLFM imagery are available elsewhere \cite{Moen2019,Gupta2019}. The vast majority of single-cell segmentation methods are designed for \emph{a posteriori} data processing and often require manual input \cite{Lugagne2020}. Many methods exist to segment yeast cells on microscope imagery, for example \cite{Wood2019,Versari2017,Dimopoulos2014,Bredies2011}. Convolutional neural networks (CNNs) are suited to the segmentation task for both \textit{E.coli} and yeast, examples include \cite{Kraus2017,Aydin2017,Lu2019,VanValen2016,Kong2020} and can perform on par or even surpass human annotation \cite{He2015,Aydin2017}. The U-Net CNN architecture enables training with relatively few annotated samples and has achieved exemplary segmentation results on various scenarios \cite{Ronneberger2015,Falk2019}.

Segmenting cells trapped in microstructures is a specific scenario for which dedicated trapping and segmentation tools are available. In the case \textit{E. coli}, which has a distinctly different morphology to yeast, the trapping devices are frequently referred to as mother machines \cite{Lugagne2020,Sauls2019}. The configuration is different to the one we consider, in particular the trap geometry is not similar to the morphology of the cells of interest and there is no risk of segmenting a trap as a cell or vice versa. Nonetheless, CNNs have been shown to perform well for this task
and U-Nets have recently been employed to segment \emph{E. coli} in mother machines  \cite{Lugagne2020,Sauls2019}.


\tdsDoneHide{check E. coli spelling (space) and first mention}


The current state-of-the-art tool for the segmentation of hydrodynamically trapped yeast cells on microfluidic chips is DISCO \cite{Bakker2017}. The pipeline extracts traps from microscope imagery and segments individual cells with conventional methods (template matching, support vector machine, active contours). Segmentation accuracy, as measured by intersect-over-union, of approximately \SI{0.7}{} is reported for the cell class. 
\section{Methodology}
\label{sec:method}
\addcontentsline{tdo}{todo}{\textbf{Methodology}}

\tdsDoneHide{review Moen2019 methods section as positive example}{
\cite{Moen2019} Moen2019 has some good example methodology. Ronneberg2015\cite{Ronneberger2015} also}

\subsection{Microscope imagery, classes and annotated dataset}
\label{ssec:data}

Individual specimen images (Fig. \ref{fig:intro_schem}) containing a single microfluidic trap and potentially some cells are extracted from larger images. The microscope automatically records an array of positions. Each exposure yields up to \si{50} traps (Fig. \ref{fig:complex}). Ideally, a single mother cell persists in a trap, with subsequent daughter cells being removed by the flow. In practice, multiple cells often accumulate around a trap. 
\begin{figure}[htbp]
	\centerline{\includegraphics{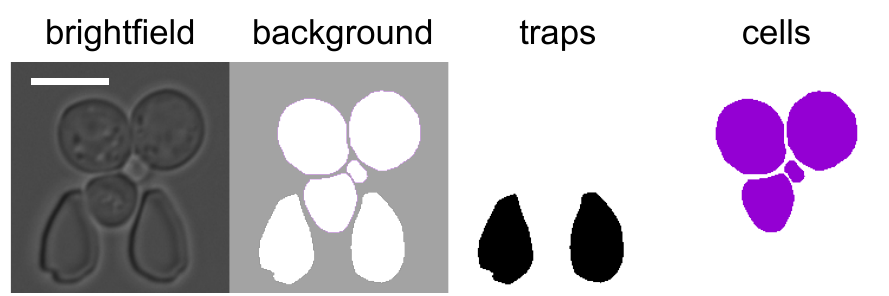}}
	\caption[$\blacksquare$ Fig. \ref{fig:classes_def}: classes (\checkmark)]{Class labels for an example of a specimen image; brightfield image (left), background label in \colorindicator{tab:bggrey}{grey}, trap labels in \colorindicator{tab:trapblack}{black} and cell labels in \colorindicator{tab:darkviolet}{violet} (left to right respectively); scale bar \SI{10}{\mu \meter}.}
	\label{fig:classes_def}
\end{figure}

We distinguish between multiple classes on the specimen images, as depicted in Fig. \ref{fig:classes_def}. The geometry of the traps is similar to the morphology of the cells, both exhibiting a roughly circular shape and similar characteristic length. To counteract traps being segmented as cells, we employ distinct classes for traps (black) and cells (violet). The background (grey) comprises all regions that are neither cell nor trap, including incomplete cells around the edge of the image or any debris on the image.

\begin{figure}[htbp]
	\centerline{\includegraphics{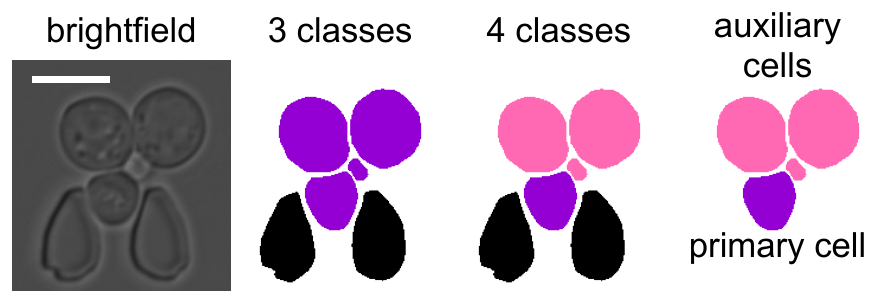}}
	\caption[$\blacksquare$ Fig. \ref{fig:classes_cells}: cell subclasses (\checkmark)]{Class labels with either one or two cell classes; 3 classes (lone cell class in \colorindicator{tab:darkviolet}{violet}), 4 classes (two cell classes), primary cells in \colorindicator{tab:darkviolet}{violet} and auxiliary cells in \colorindicator{tab:hotpink}{light violet} (left to right respectively); scale bar \SI{10}{\mu \meter}.}
	\label{fig:classes_cells}
\end{figure}

The cell class can be further separated into primary and auxiliary cells along the lines of expected on-chip stay (Fig. \ref{fig:classes_cells}). We define primary cells as those firmly caught in the trap microstructures, which by design are expected to remain on chip for the duration of an experiment. The auxiliary cells, for example daughters of primary mother cells, should be hydrodynamically washed out of the chip by the continuous media flow (top to bottom, Fig. \ref{fig:classes_cells}) to avoid clogging of the chip. This separation may be advantageous to optimise primary cell segmentation adjacent to traps, as well as for further downstream data analysis, for example of microfluidic chip performance for mother cell retention and daughter removal. We annotated four classes on each image. Three class labels were automatically derived by joining the two cell classes.
\tdsDoneHide{decide: $\rightarrow$no: add AF interfacial class?}{add Andr\'e's interface class as it further improves stuff or leave for follow up pipeline? $\rightarrow$ decide after making pipeline outline of increment beyond this}
\tdsDoneHide{change colours to violet/plum for 3 cl? $\rightarrow$ no (time)!}{}

\begin{figure}[htbp]
	\centerline{\includegraphics{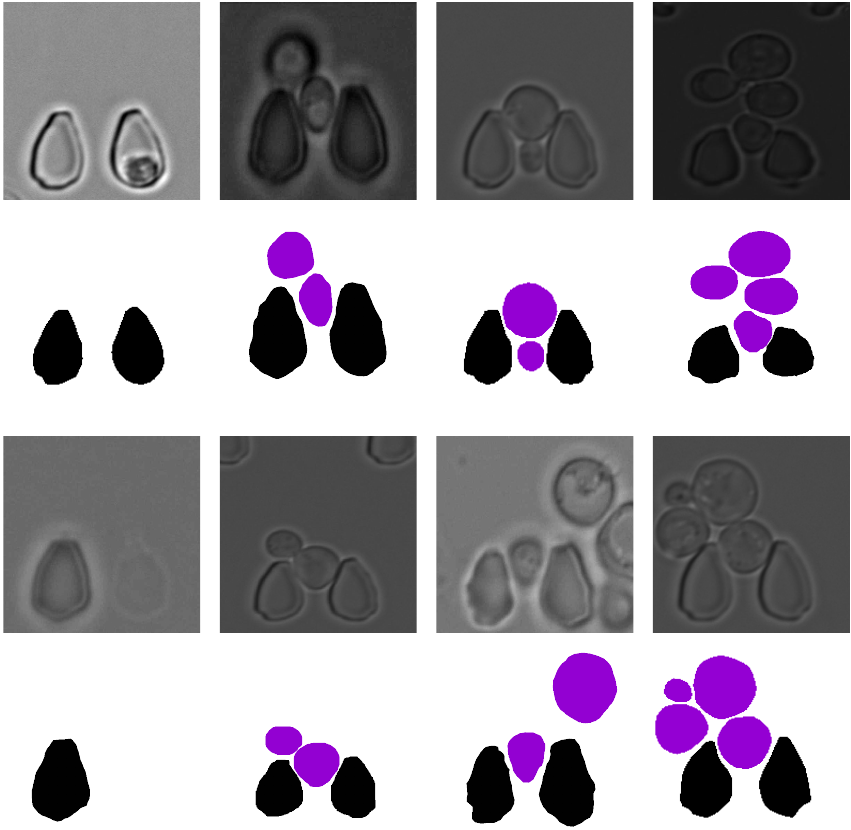}}
	\caption[$\blacksquare$ Fig. \ref{fig:tradata}: training data (\checkmark)]{Characteristic selection of specimen images and corresponding labels from the training dataset, including empty or single trap structures, trapped single cells (with single daughter adjacent) and multiple trapped cells; traps in \colorindicator{tab:trapblack}{black}, cells in \colorindicator{tab:darkviolet}{violet} and transparent background.}
	\label{fig:tradata}
\end{figure}

Segmentation models were trained, validated and tested on an annotated set of $633$ specimen images taken from numerous experiments. Examples are shown in Fig. \ref{fig:tradata}. The training  data comprised \si{487} specimen images, with a further \si{65} and \si{81} reserved for validation and testing respectively.  We selected images to include a balance of the common yeast-trap configurations: 1) empty traps, 2) single cells (with daughter) and 3) multiple cells. Slight variations in trap fabrication, debris and contamination, focal shift, illumination levels and yeast morphology were included. 
We included a balance of empty and filled traps to prevent poor segmentation results  when no cells are present. Further possible scenarios or strong variations were omitted, such as other trap design geometries, other model organisms and significant focal shift. 

\tdsDoneHide{resolve: dataset solution $\rightarrow$ split:  train+val+test}{split dataset into three parts, trian, val, test; with additional application test data (unlabelled) for further qualitative analysis of generalisation / robustness}
\tdsDoneHide{resolve dataset issues}{add test dataset as a similar set to the training data, not randomly chosen as both need to be balanced with regard to the highly unbalanced occurances (ie empty traps etc. need to be included in both)}
\tdsDoneHide{dataset split valAF $\rightarrow$ val+test  $\rightarrow$ train+val+test; +appl.}
{CR: no val dont explain, highly unbalanced dataset, we chose train and test sets to be balanced, too few samples to do it randomly}
\tdsDoneHide{remedy class labels redundancy? here or in subclasses?}{\P contd as follows: We labelled each image with four classes, two of which denoting the primary and auxiliary cell classes. Three class labels were automatically derived by joining the two cell classes.}

%

\begin{figure}[htbp]
	\centerline{\includegraphics{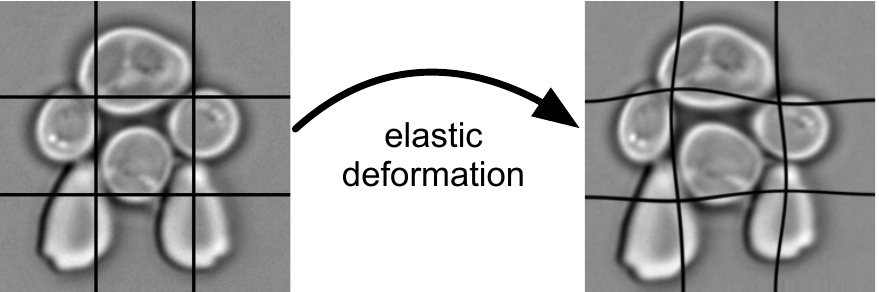}}
	\caption[$\blacksquare$ Fig. \ref{fig:augmentation}: data augmentation (\checkmark)]{Example of training sample augmentation by elastic deformation.
	} 
	\label{fig:augmentation}
\end{figure}

We augmented the training dataset by elastically deforming each specimen image and label pair once randomly \cite{Ronneberger2015,Simard2003}. An example deformation is shown in Fig. \ref{fig:augmentation}. The displacement of a $3\times3$ grid was drawn from a normal distribution $\mathcal{N}(0;1)$ and smoothed by a gaussian filter with standard deviation of 3 pixels. Pixel displacement is interpolated (bicubic). 
\tdsDoneHide{'Drop-out layers: we didnt use them}{at the end of the contracting path perform further implicit data augmentation' did we use them in these specific models, I know we discussed them}

\tdsDoneHide{elastic deformation Simard2003? }{first paper from CR 20.06.2020}
\tdsDoneHide{no; decide: omit augmentation figure?}{}

\tdsDoneHide{no: add example application data \P?}{or just reference a bit in the introduction/background/explanation of the applciation data earlier on?!extend to time-lapse recording? Time-lapse recordings and enable.}

\subsection{Network architectures, training and evaluation }
\label{ssec:approach}
\addcontentsline{tdo}{todo}{\textbf{Methodology: Approach}}

\tdsDoneHide{decide on which models to present and describe those}{Model \#4 and \#1, with wloss=0; wloss=1; 3 classes, 4 classes, with/without interfacial class,}
\tdsDoneHide{architecture A original and B the modified version! CW the following differences betwene the two, omit the long description.}{}

\begin{figure*}[htbp]
	\centerline{\includegraphics[width=1\textwidth]{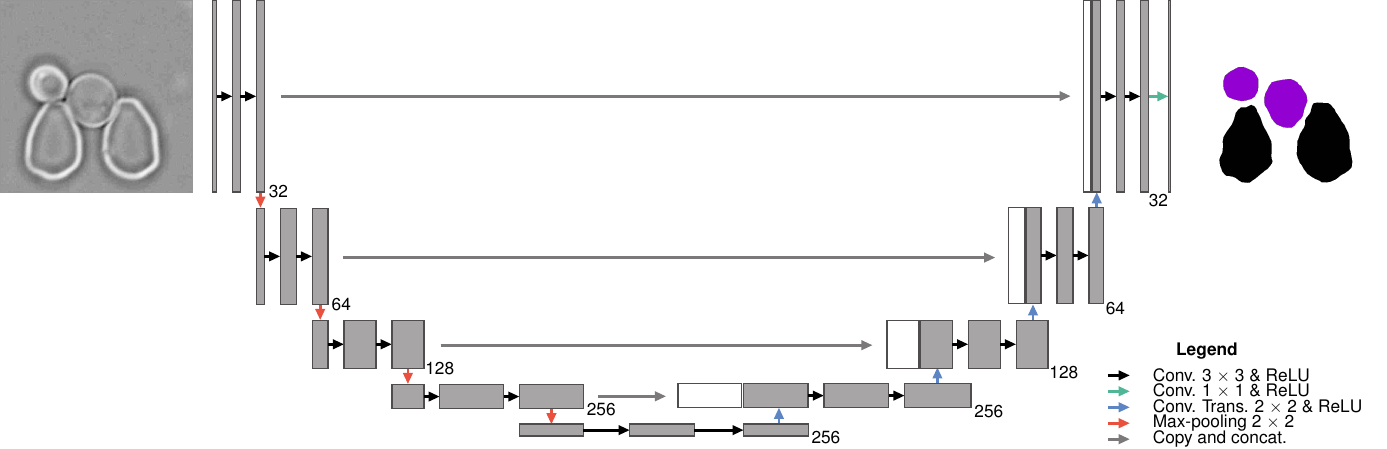}}
	\caption[$\blacksquare$ Fig. \ref{fig:architecture2} unet architecture double column (\checkmark)]{Architecture of Model B, with brightfield input and a segmentation prediction as output after passing through $3\times3$ convolutions in \colorindicator{tab:trapblack}{black}, max pooling of the encoding branch in \colorindicator{tab:TUred9a}{red}, long connections between encoder and decoder blocks in \colorindicator{tab:tud0c}{grey}, transpose covolutions in the decoding branch in \colorindicator{tab:TUblue1a}{blue} and the final convolution in \colorindicator{tab:TUgreen3a}{green}; block width and numerals indicate the number of feature maps in each block, block height indicates the XY-resolution starting at $128 \times 128$.}
	\label{fig:architecture2}
\end{figure*}

We adapted the architecture of the original U-Net \cite{Ronneberger2015} to segment our cell, trap, and background classes. Here we present the two models, outlined in Table \ref{tab:architecture}. Model A is similar to the original U-Net. We reduced the size of the network in model B, to decrease the computational cost of training and inference. The structure of model B is shown in Fig. \ref{fig:architecture2}.

The input into the networks is a brightfield image of $128 \times 128$ pixels, which is halved in each level of the encoding path. Model B utilises $32$ filters in the first layer, in contrast to $64$ for model A. The feature dimensions are doubled in each encoder stage until reaching $256$ for model B in both the fourth level and in the bottleneck. For model A the feature dimensions continue to double up to $1024$ in the bottleneck.

The decoder path reduces the number of feature dimensions and doubles the XY-dimension in each stage. The input to each of the four decoder stages is a concatenation in the feature channel dimension of the upsampled output of the previous level and the output of the corresponding encoder block. The output has the dimension of the corresponding encoder level. The final layer maps the $64$ or $32$ feature channels to the desired number of segmentation classes by $1 \times 1$ convolution.

\begin{table}[htbp]
	\caption[$\boxminus$ Table \ref{tab:architecture}: model architecture (\checkmark)]{Overview of the model A and B architectures for $K$ classes.}
	\centering
	\begin{tabular}{ c c c c c c} 
		\toprule
		\multicolumn{1}{p{1.2cm}}{\centering Architecture\\ \ } &
		\multicolumn{1}{p{1.2cm}}{\centering Input}	&
		\multicolumn{1}{p{1.2cm}}{\centering Depth} &
		\multicolumn{1}{p{1.2cm}}{\centering Feature \\Channels } &
		\multicolumn{1}{p{1.2cm}}{\centering Output} \\ 
		\midrule
		A  	& $128^2$ 	& $1$	& $64$ 		& $128^2 \times K$  \\
		&  			& $2$	& $128$ 	&  	\\
		&  			& $3$	& $256$ 	&  	\\
		&  			& $4$	& $512$  	& 	\\
		&  			& $5$	& $1024$ 	&  	\\
		\midrule
		B  	& $128^2$ 	& $1$	& $32$ 		& $128^2 \times K$  \\
		&  			& $2$	& $64$ 		&  	\\
		&  			& $3$	& $128$ 	&  	\\
		&  			& $4$	& $256$  	& 	\\
		&  			& $5$	& $256$ 	&  	\\
		\bottomrule\\
	\end{tabular}
	\label{tab:architecture}
\end{table}


\tdsDoneHide{choose which architecture table}{}
		


\tdsDoneHide{fig architecture implement required changes}{$\bullet$ change the height of the blocks\\
	$\bullet$ decoder block widths are wrong\\
	$\bullet$ see handwritten notes\\
	$\bullet$ make the input output images bigger}

\begin{figure}[htbp]
	\centerline{\includegraphics{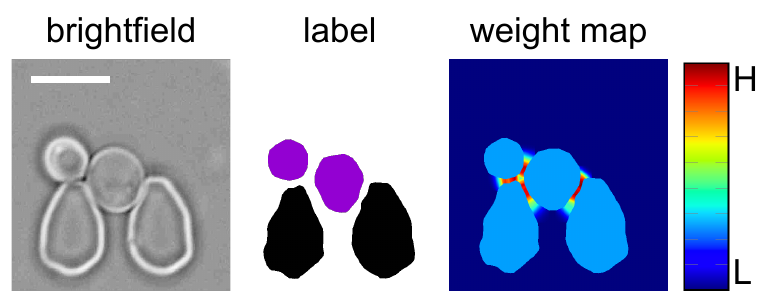}}
	\caption[$\blacksquare$ Fig. \ref{fig:weight_map}: weight map figure \checkmark]{Weight map, weights from blue background (low) to most important regions between cells and traps in red (high), with traps and cells themselves twice the weight of the background (light blue) for a given example brightfield image and corresponding annotated label; scale bar \SI{10}{\mu \meter}.}
	\label{fig:weight_map}
\end{figure}

We chose the loss function $\lambda$ with the aim of avoiding a bias towards the background class due to the class imbalance. We employed the sum of S{\o}rensen-Dice loss $\lambda_s$ and pixel-wise weighted cross-entropy loss $\lambda_e$ \cite{Sudre2017,Ronneberger2015}. For imagery of resolution $M \times N$, pixel labels $y$, prediction $\hat{y}$ and $K$ classes the loss function components are
\begin{equation*}
 \lambda_{s} = 1 -  \frac
 {2 \sum\limits_{k=1}^{K} \sum\limits_{m=1}^{M} \sum\limits_{n=1}^{N} y_{m,n,k} \ \hat{y}_{m,n,k} +\gamma} 
 {\sum\limits_{k=1}^{K} \sum\limits_{m=1}^{M} \sum\limits_{n=1}^{N} (y_{m,n,k} + \hat{y}_{m,n,k}) +\gamma}
\label{eqn:Ls} 
\end{equation*}
\tdsDoneHide{loss: 1/K in loss functions?}{} 
and 
\tdsDoneHide{loss: smoothness}{dice paper for smoothness factor and decide whether to put it in, epsilon in dice paper Shen and Li both have loss}
\tdsDoneHide{either: double check nomenclature $\mathcal{L}$ or $\lambda$ with CW}{}
\tdsDoneHide{Dice paper, CR 27.06.2020}{}
\begin{equation} 
\lambda_{e} = \frac{1}{K M N} \sum_{m=1}^{M} \sum_{n=1}^{N} w_{m,n} \sum_{k=1}^{K} -y_{m,n,k} \log \hat{y}_{m,n,k}
\label{eqn:Le}
\end{equation} 
with $w$ pixel-wise weights and $\gamma = 1$ for smoothing. To better discriminate between single instances of the cell and trap objects, the weights are high in the particularly important regions between cells, and between traps and cells. An example is depicted in Fig. \ref{fig:weight_map}. Weight maps are calculated based on a distance transform of the data labels \cite{Ronneberger2015}. Some networks were trained with uniform weights, effectively without weighting. 
\tdsDoneHide{decide: add weight map image?CR or AF}{CR has the nicer weight map images load form CR dataset with torch.load filename.numpy[]\\ AF also has some decent ones and they should be accessible, the iamge is the same one as the dfromation one which is nice}
\tdsDoneHide{not found: add reference for CE+DICE?}{CR has a reference for thsi being the best method}

\tdsDoneHide{weighted loss weight map, follow up Long2014 reference?}{weighted loss due to class mismatch Long2014? Emphasise the cell class or the primary cell class?!}
\tdsDoneHide{decide detail on weight distance transform?}{add description of weight map / distance transform?}

We implemented the described network architectures in TensorFlow \SI{2.0}{}, and minimised the loss function with the Adam optimiser \cite{Kingma2015}. The learning rate was set to $0.001$, the first and second order momentum moving average factors to $0.9$ and $0.999$ respectively. The learnable parameters were initialised with the Kaiming He scheme \cite{He2015}. We manually tuned the hyperameters on the validation data. The trained networks were selected for minimum validation loss, typically found after \SI{30} epochs, with a batch size of \SI{8}{}. Computations were performed with two Intel  Xeon  Gold  (Skylake)  6148 2.4GHz  CPUs (without  hyperthreading) and a GeForce RTX 2080 Ti GPU with 11 GB VRAM.
\tdsDoneHide{Kaiming He NORMALISATION?}{}
\tdsDoneHide{add GPU information!!!}{}
\tdsDoneHide{add information to computational setup from AF? }{capable of running on personal laptop in addition to on our servers}

\tdsDoneHide{add following hyperparameters to implementation}{
$\bullet$ training rate\\
$\bullet$ initialisation Kaiming He\\
$\bullet$ learning rate: Adam optimizer that starts with a learning rate but then takes on an adaptive learning rate\\
$\bullet$ other hyper-parameters?
}

\tdsDoneHide{split validation set into validation and training set!}{}
\tdsDoneHide{mini batch size $\rightarrow$ 8}{}
\tdsDoneHide{add Adam optimizer information}{}
\tdsDoneHide{epoches / training termination contains all the data}{}
\tdsDoneHide{mention tensorflow (here)}{}
\tdsDoneHide{spelling: hyperparameter}{}

\tdsDoneHide{add pre-processing (not trap extraction)}{only add the essential steps, not trap extraction}
\tdsDoneHide{add normalisation (not He initialisation)}{Where does normalisation go?}
\tdsDoneHide{no?: add one hot encoding?}{Could be here or in implementation \P}
The specimen images were pre-processed before being fed into the U-Net. We downsampled the input to a resolution of $128 \times 128$, which was deemed the best trade-off between image quality, segmentation performance and speed after preliminary evaluation on the validation set. We normalised each input to a mean of zero and a variance of one, to counteract the variation in illumination and histograms across microscope recordings.

To quantitatively analyse the performance of the trained networks, we employ the mean Jaccard index (intersecetion-over-union) and the S{\o}rensen-Dice coefficient. Given the ground truth $\mathbf{Y}$ and prediction $\mathbf{\hat{Y}}$, the S{\o}rensen-Dice coefficient is
\begin{equation*}
S(\mathbf{Y},\mathbf{\hat{Y}}) = \frac{2 |\mathbf{Y} \cap \mathbf{\hat{Y}}|}{|\mathbf{Y}| + |\mathbf{\hat{Y}}|}.
\end{equation*}
The Jaccard index for class $k$ is
\begin{equation*}
	J_k(\mathbf{Y_k},\mathbf{\hat{Y}_k}) = \frac{|\mathbf{Y_k} \cap \mathbf{\hat{Y}_k}|}{|\mathbf{Y_k} \cup \mathbf{\hat{Y}_k}|}.
\end{equation*}
We employ the mean intersection-over-union over the $K$ classes $\bar{J} =\frac{1}{K} \sum_{k = 1}^{K} J_k$ to counteract the background class imbalance. We also consider the Jaccard index for the cell class alone ($J_c$), as segmentation of the cells is of particular importance for the application in image cytometry. 

\tdsDoneHide{mention cell class IoU? going forward IoU of cell only}{only the cell is relevant for the measurement task}

Evaluating segmentation metrics quantitatively is limited to the available labelled data. To further qualitatively analyse how robust the models are toward variations in the segmentation imagery, we also verified segmentation performance qualitatively on additional data. The  application example recordings  stem from different experiments, include time-lapse recordings, as well as the fluorescent channel. This allowed us to also trial time-series measurements of single-cell fluorescence. 
\tdsDoneHide{extend to time-lapse recording?}{Time-lapse recordings and enable.}
\tdsDoneHide{extend to qualitative analysis based on unlabelled data}{}
\tdsDoneHide{metrics $\rightarrow$ quantitative analysis given labelled data}{}
\tdsDoneHide{unlabelled data tests for further qualitative analysis of robustness}{}

\subsection{Data acquisition setup}
\label{ssec:acquisition}
\addcontentsline{tdo}{todo}{\textbf{Methodology: Data Acquisition}}

Microfluidic chips confined yeast cells to the focal plane of the microscope and ensured an environment conducive to yeast growth. Our microfluidic chip is depicted in Fig. \ref{fig:complex}. Continuous media flow hydrodynamically trapped live yeast cells in microstructures. The cells were laterally constrained in XY by Polydimethylsiloxane (PDMS) microstructures and axially in Z by the cover slip and PDMS ceiling. The space between the cover slip and the PDMS ceiling is on the order of a cell diameter to facilitate uniform focus of the cells. A temperature of \SI{30}{\celsius} and the flow of yeast growth media enables yeast to grow for prolonged periods and over multiple cell-cycles.
\tdsDoneHide{remove some redundancy}{}
\tdsDoneHide{the media flow feeds the cell traps.}{}

A computer controlled microscope (Nikon Eclipe Ti with XYZ stage; $\mu$Manager; 60x objective) recorded time-lapse brightfield (transmitted light) and fluorescent imagery of the budding yeast cells every \SI{10}{\min}. Multiple lateral and axial positions are recorded sequentially at each timestep. A CoolLED pE-100 and Lumencor SpectraX light engine illuminated the brightfield and fluorescent channel images respectively, which were recorded with a ORCA Flash 4.0 (Hamamatsu) camera. 
\tdsDoneHide{first sentence bracket improve}{first sentence remove or significantly reduce bracket content}
\tdsDoneHide{illumination brightfield source}{}

\section{Results}
\label{sec:results}
\addcontentsline{tdo}{todo}{\textbf{Results}} 
\subsection{Performance of trained networks}
\label{ssec:res_test}

\begin{figure}[htbp]
	\centerline{\includegraphics{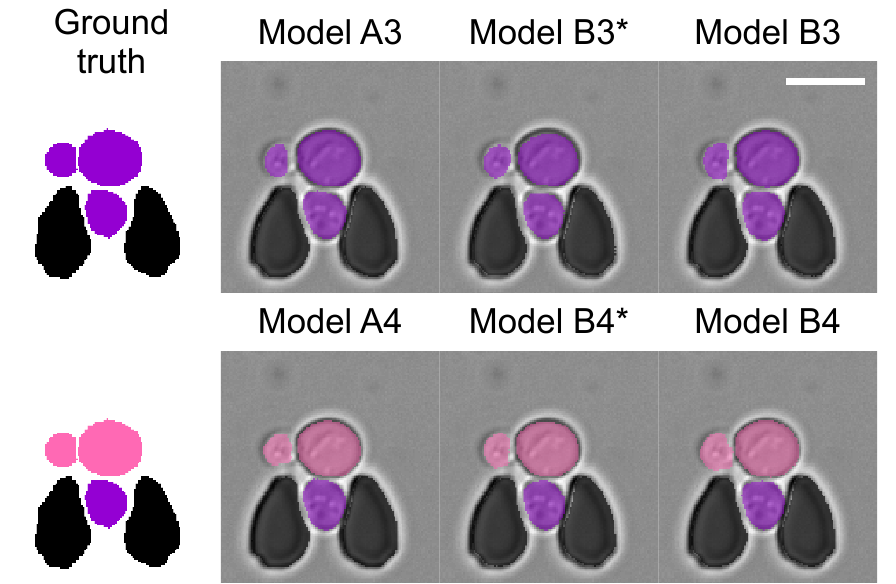}}
	\caption[$\blacksquare$ Fig. \ref{fig:res_val}:  segmentation examples \checkmark]{Example segmentation for various networks. Transparent background, traps in \colorindicator{tab:trapblack}{black} and cells in \colorindicator{tab:darkviolet}{violet} for three classes and primary cells in case of four classes with \colorindicator{tab:hotpink}{light violet} auxiliary cells; scale bar \SI{10}{\mu \meter}.}
	\label{fig:res_val}
\end{figure}

A sample segmentation for various trialled networks is shown in Fig. \ref{fig:res_val}. The background (transparent), traps (black) and all cells (violet) are segmented as the correct classes with slight variations in the cell and trap contours. In the case of four classes, the primary (violet) and auxiliary (light violet) cells are successfully discriminated. 
\tdsDoneHide{discussion: add mention, 4cl a step toward instance detection/seg?}{}

The results of the experiments on the test data are summarised in Table \ref{tab:test}. We trained two separate architectures for three and four class semantic segmentation, as well as with two different loss functions (trained models are available from the authors). The network variants performed similarly well on the test data as measured by the S{\o}rensen-Dice $S$ and mean Jaccard index $\bar{J}$ metrics. The best model achieved $S = 0.96$ and $\bar{J} = 0.89$ for three classes trained with the weighted loss function. The pixel-wise weighted loss function outperformed the unweighted variant in every case. 

\begin{table}[htbp]
	\centering
			\caption[$\boxminus$ Table \ref{tab:test}: test metrics \checkmark]{Network test results. $\lambda_{e}^{*}$ indicates uniform weighted cross-entropy loss (Eqn.\ref{eqn:Le}, $w_{m,n} = 1$ for all $m,n$).}
		\begin{tabular}{ c c c c c c} 
			\toprule
			\multicolumn{1}{p{1.2cm}}{\centering Architecture\\ \ }		& \multicolumn{1}{p{1.2cm}}{\centering Number \\ Classes }	& \multicolumn{1}{p{1.2cm}}{\centering Loss\\ Function}		&
			\multicolumn{1}{p{1.2cm}}{\centering S{\o}rensen\\$S$ } 	& \multicolumn{1}{p{1.2cm}}{\centering Jaccard \\$\bar{J}$ } \\ 
			\midrule
			A  & $3$ & $\lambda_{s} + \lambda_{e}^*$	& $0.9594$ 	& $0.8779$  	\\
			A  & $3$ & $\lambda_{s} + \lambda_{e}$ 		& $\mathbf{0.9637}$  & $\mathbf{0.8870}$  \\
			A  & $4$ & $\lambda_{s} + \lambda_{e}^*$ 	& $0.9540$ 	& $0.8212$ 	\\
			A  & $4$ & $\lambda_{s} + \lambda_{e}$ 		& $0.9554$ 	& 0.8349  	\\
			\midrule
			B  & $3$ & $\lambda_{s} + \lambda_{e}^*$ 	& $0.9585$ 	& $0.8783$  \\
			B  & $3$ & $\lambda_{s} + \lambda_{e}$ 		& $\mathbf{0.9626}$ & $\mathbf{0.8839}$  \\
			B  & $4$ & $\lambda_{s} + \lambda_{e}^*$ 	& $0.9538$ 	& $0.8301$  \\
			B  & $4$ & $\lambda_{s} + \lambda_{e}$ 		& $0.9556$ 	& $0.8351$ 	\\
			\bottomrule\\
		\end{tabular}
	\label{tab:test}
\end{table}

\tdsDoneHide{no:change Jaccard and Sorensen terms?!}{}
\tdsDoneHide{later:add cell class IoU}{potentially add accuracy for discussion too? accuracy can also be added to the text}
\tdsDoneHide{post submission: giga flops for runtime analysis}{giga flops for runtime analysis, page 9 on DETR paper}
\tdsDoneHide{done go into: CE+DICE problem, CE is much larger}{common practice to balance empirically, we didn't really do it.}

The model architecture has little effect on the evaluation metrics. The performance of the large network A, akin to the original U-Net \cite{Ronneberger2015}, was effectively conserved in the significantly smaller network B. We  reduced the number of parameters from approximately $31 \times 10^6$ to $5 \times 10^6$ and the floating point operations (FLOPs) for a forward pass reduced fourfold from $24 \times 10^9$ to $6 \times 10^9$. We timed network B3 to segment a specimen image in $\SI{9}{\milli\second}$, approximately $3$ times faster than the larger network.
\tdsDoneHide{add GPU details?}{}
\tdsDoneHide{omit mention of computational overhead}{ as cause for 6-fold to 4-fold  3-fold?}
\tds[disable]{add Parameter number and run times?}{ to table if it is needed for space}
\tdsDoneHide{double check in pytorch}{}

\tdsDoneHide{add introductory sentence to this paragrpah?}{}
Increasing the number of classes had a small negative effect on the overall Dice loss $S$ and a larger effect on the mean Jaccard index $\bar{J}$ which reduced from $0.89$ to $0.83$. One reason for the stronger decrease in $\bar{J}$ is that the most challenging class, the cells, contribute more strongly to the mean in this case. Beyond this, an additional source of error is introduced by misclassification between the cell classes.

\tdsDoneHide{add mention of class mixup }{reducing the mIoU in addition to effectively more weight on the class}
\subsection{Examples of segmentation for test and fluorescence  data}
\label{sssec:res_bestsamples}
\begin{figure}[htbp]
	\centerline{\includegraphics{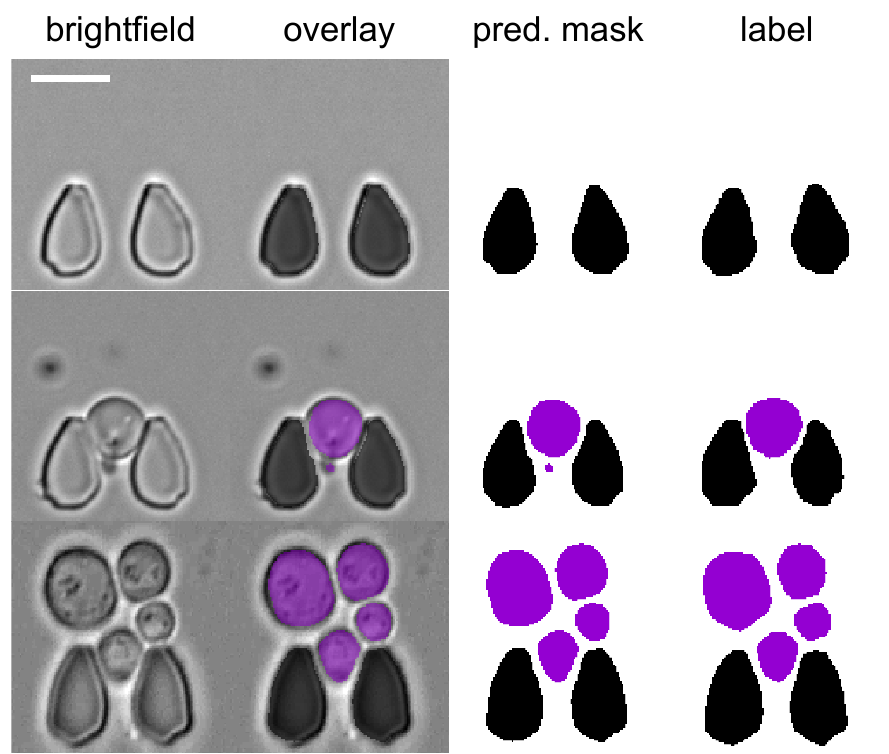}} 
	\caption[$\blacksquare$ Fig. \ref{fig:res_test}: test segmentation examples (\checkmark)]{Example of different scenarios segmented with B3: an empty trap (top row), a single trapped cell with daughter (middle row) and multiple cells; columns are brightfield, an overlay of the prediction, the prediction mask and the ground truth label  (left to right respectively). Transparent background, colours indicate traps in \colorindicator{tab:trapblack}{black} and cells in \colorindicator{tab:darkviolet}{violet}; scale bar \SI{10}{\mu \meter}.}
	\label{fig:res_test}
\end{figure}

Examples of segmentation predictions made by network B3 for three typical scenarios are given in Fig. \ref{fig:res_test}. Balancing of the training dataset to include enough empty traps ensured that these are not mistakenly detected as cells (top row). Detection of trapped single cells, potentially with an attached daughter, is the standard design case (middle row). When multiple cells are present it is important that these are segmented individually to facilitate simple instance detection of each individual cell object. Together, the introduction of multiple classes and of the pixel-wise weighted loss function facilitated individually segmenting each cell and discerning these form the traps.

\tdsDoneHide{Results: add touching cells?}{ add some data on the touching cells issue?}
\tdsDoneHide{where done? add cell class IoU mention to results}{}

\tdsDoneHide{add bit on weighted loss function is better for this?}{does this require some artowrk as exmaple of how weighted loss is better? the improvement is larger than the marginal gain in the validation metric!}
\tdsDoneHide{discriminate between traps and cells}{}
\tdsDoneHide{yes,decide: add fluoresence imagery here?}{}
\tdsDoneHide{no,decide: add time-lapse traces?}{}

\begin{figure}[htbp]
	\centerline{\includegraphics{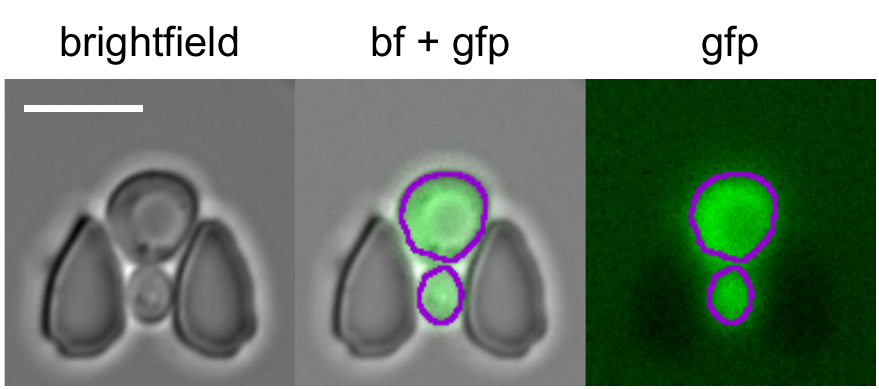}} 
	\caption[$\blacksquare$ Fig. \ref{fig:fluoro}: fluoro seg \checkmark]{Sample segmentation  on the brightfield channel (left) for fluorescence measurement on the green fluorescent protein (GFP) channel (right) and an overlay of the two (middle); scale bar \SI{10}{\mu \meter}.}
	\label{fig:fluoro}
\end{figure}

The proposed method is designed to deliver segmentation masks for subsequent single-cell fluorescence measurements. An example of this on the unlabelled data is given in Fig. \ref{fig:fluoro}. Our method predicts a segmentation mask for each cell based on the brightfield image alone. The mask contour is depicted  in violet on both the green fluorescent channel (GFP), and on the overlay of green fluorescence and brightfield (middle). 
\tdsDoneHide{add \P \ on fluorescence and application test resul}{}
\tdsCamera{probably not here...mention runtime here?}{}
\tdsCamera{add scale fluoro image?}{}


\tdsDoneHide{what to show for application tests}{
	$\checkmark$ show predicted mask on a fluorescence channel image?!\\
	$\checkmark$ only first: show segmentation examples only or go to area plots and fluorescence?\\}

\tdsDoneHide{decide: time-lapse fluoro in/out?}{decide how far to go with these results in direction of time-lapse fluorescent imagery}

\subsection{Comparison with the state-of-the-art} 
\label{ssec:rescomp}
\addcontentsline{tdo}{todo}{\textbf{Results: \nameref{ssec:rescomp}}}
Accurate segmentation of the cells is particularly important for the measurement of cell morphology or fluorescence. We compare the cell class Jaccard index for model B3 with the reported performance of the state-of-the-art trapped yeast segmentation pipeline DISCO \cite{Bakker2017} in Table \ref{tab:comparison}. DISCO was reportedly tested on three separate datasets with similar traps to ours; we chose the best reported results for comparison. DISCO ($J_c  \sim \SI{0.7}{}$) outperforms the previous methods \cite{Bredies2011,Dimopoulos2014,Versari2017} ($J_c \leq \SI{0.6}{}$) and our proposed method achieves a significantly better Jaccard index $J_c = \SI{0.82}{}$ for the cell class.

\tdsDoneHide{work microstructured in?}{}

\begin{table}[htbp]
	\centering
			\caption[$\boxminus$ Table \ref{tab:comparison}: comparison cell IoU \checkmark]{Comparison of cell class intersection-over-union $J_c$ with reported state-of-the-art \cite{Bakker2017}. }
		\begin{tabular}{ l r} 
			\toprule
			Method 									&	 Cell Class $J_c$ \\
			\midrule
			DISCO \cite{Bakker2017} 	 			& $\sim0.70$ \\
			DISCO + CellSerpent \cite{Bakker2017}	& $\sim0.60$\\
			DISCO + CellX \cite{Bakker2017}			& $\sim0.55$\\
			DISCO + CellStar \cite{Bakker2017}		& $\sim0.40$\\
			\textit{Proposed method} (B3)  			& $0.82$ \\
			\bottomrule\\
		\end{tabular}
	\label{tab:comparison}
\end{table}

\tdsCamera{change table caption style? not neccessary}{}
\tdsCamera{shorten caption and move most of it to text}{Comparison of cell segmentation performance with the best performing trapped yeast segmentation pipeline available \cite{Bakker2017}. Trap shapes differed between our analysis and that of DISCO and different datasets were employed. For as fair a comparison as possible, we chose the best reported segmentation performance for DISCO out of three different datasets.}
\tdsCamera{no: add iou cell class for other models too? egA4}{}

We tested the runtime of the proposed segmentation method in a rudimentary pipeline that takes large microscope images as its input, detects traps and segments each cell. A comparison with reported runtimes for DISCO \cite{Bakker2017} is presented in Table \ref{tab:runtimes}. Traps were detected by cross-correlation template matching, similar to that of DISCO \cite{Bakker2017}. We processed large images ($2048 \times 2048$ pixels) at $79$ successive timepoints, each with $37$ traps and across $5$ focal planes. In total, we processed \SI{6.2}{GB} of data in under \SI{4}{\minute}, or approximately \SI{20}{\milli\second} per specimen image on average. In comparison, DISCO reportedly requires \SI{\sim 1}{\second} per specimen image on average ($133$ large images, $512 \times 512$ pixels, with $45$ traps each in at least \SI{100}{\minute}). We indicate the FLOPs required for a forward pass of our network to facilitate future comparisons in Section \ref{ssec:res_test}. 

\begin{table}[htbp]
	\centering
			\caption[$\boxminus$ Table \ref{tab:runtimes}: comparison runtime \checkmark]{Comparison of runtimes reported for DISCO \cite{Bakker2017} and model B.}
		\begin{tabular}{ l r} 
			\toprule
			Method 								& runtime per specimen \\
			\midrule
			DISCO \cite{Bakker2017}  			& $\sim\SI{1300}{\milli \second}$ \\
			DISCO + CellX \cite{Bakker2017}		& $\sim\SI{1000}{\milli\second}$\\
			\textit{Proposed method} (B3)  		& $\sim \SI{20}{\milli \second}$ \\
			\bottomrule\\
		\end{tabular}
	\label{tab:runtimes}
\end{table}
\tdsDoneHide{add mention of flops better for analysis (in results?)}{}






\section{Discussion}
\label{sec:discussion}
\addcontentsline{tdo}{todo}{\textbf{Discussion}}


\tdsDoneHide{no; decide: mingle Discussion and Results?}{}

\subsection{Analysis of multiclass segmentation performance} 
\label{ssec:qualitative}

The trained networks achieve a S{\o}rensen-Dice coefficient $S$ of $0.96$ and mean Jaccard index $\hat{J}$ of $0.89$, approaching human level annotation. They are able to generalise and compensate some limitations in the annotated data. For example, in one particular edge-case (Fig. \ref{fig:res_test}, central row) the network detects the presence of a very small daughter cell. The daughter cell in question is so small, that it was not annotated in the test data; such small cells were generally omitted in labelling, due to potential issues such as their position and movement relative to the focal plane. In contrast to human labelling, the model prediction is reproducible for a given image and may exhibit less variation between similar images, which is deemed advantageous when extracting information across multiple similar time-points.
%
%

\tdsDoneHide{take focus off human lvl, merely example in paragrpah}{}


One potential source of error is mistaking a trap for a cell, which would be particularly detrimental in the analysis of population heterogeneity. The multiclass approach was successful at counteracting this by discerning the trap and cell classes. To a somewhat lesser extent primary and auxiliary cells can be discriminated. Another potential source of error occurs when the semantic segmentation prediction for objects touch, whereby they may be identified as a single object. In addition to multiclass segmentation alleviating this problem between cells and traps, the weighted cross-entropy loss function successfully counteracted this for cells of the same class.

\tdsDoneHide{check where cell-trap, cell-cell; unify line}{}
\tdsDoneHide{important to exapand on multiclass segmentation here}{}

\tdsDoneHide{include weaker dataset mention as result of paper}{}
\tdsDoneHide{training paragraph on advantages of unet?}{}
\tdsDoneHide{(careful) change training qualitative analysis to focus on dataset}{, works for trianing however is limited in scope --> extend outlook}
Segmentation of cells is not only the basis for measuring morphological properties such as shape, or for measuring cell fluorescence, but also for identifying each instance of cell objects and subsequently tracking them through time. Therefore the Jaccard index for the cell class $J_c$ is of particular importance for the time-lapse image cytometry application. The accuracy achieved in this study, which significantly surpasses previous approaches for the specific application scenario, promises to enable the generation of a novel imaging pipeline with decreased segmentation uncertainty, increased temporal tracking and subsequent experimental data yield.

\subsection{Limitations, outlook and future potential} 
\label{ssec:quantitative}
The presented networks and annotated dataset are designed for a specific microfluidic configuration and trap geometry. They are relatively robust and fulfil their intended purpose. For example in the middle and bottom row of Fig. \ref{fig:res_test}, debris is not falsely predicted to be a cell, even when of similar size and shape as the cells. Nonetheless, the training dataset is limited in scope to the specific scenario outlined in \ref{ssec:data} with a single trap geometry, limited range in focal depth or cell morphology, for example. To further generalise the applicability of the models, the training dataset could in future be extended to include a more diverse set of trapped yeast experiments, for example more axial positions, or even further to include trapped droplets for \emph{in vitro} experiments or completely different configurations. It remains to be seen how the existing architecture will perform with increasingly complex datasets and how these may be optimised for fast, accurate and precise segmentation in a wide variety of configurations and microstructured environments for single-cell TLFM. 
\tdsDoneHide{decide: really express the limitations so strongly here?}{}
\tdsDoneHide{link debris to trap-cell, cell-cell discern multiclass?}{}
\tdsDoneHide{add the limitations, loop back ot the dataset}{}
\tdsDoneHide{add commentary on 4cl stuff, check resutls for info}{}

%

The runtime of existing hydrodynamically trapped yeast segmentation pipelines, such as DISCO \cite{Bakker2017}, is generally prohibitive for online monitoring during experimentation. Utilising the reduced architecture and computational parallelisation, we segmented single specimen images within \SI{20}{\milli \second} on average. This fast segmentation capability makes online time-lapse image cytometry of hydrodynamically trapped yeast cells feasible for a typical experiment comprised of approximately \SI{1000}{} traps across \SI{20}{} positions and imaged once every \SI{10}{\min}. This capability is enabling for long-term closed-loop optimal experimental design and promises to increase the information content yield of each experiment \cite{Prangemeier2018,Gomez2019}.
\tdsDoneHide{add microstructured environment to start of this...}{}
\tdsDoneHide{yes: doneadd OED mention here?}{}
\tdsDoneHide{yes, removedredundancy: \nameref{sssec:outlook_pipeline}}{}

\tdsDoneHide{swap order with previous \P?}{}
\tdsDoneHide{is in BG: check loop ie that tracking mentioned in intro/background}{}

\tdsDoneHide{loop back to tracking and data yield?}{}

\tdsDoneHide{add mention of cell class IoU to the Results section}{}
\tdsDoneHide{calcl IoU for primary cell class?}{irrelevant, somewhat lower in general for 4cl}
\tdsDoneHide{Bakker seems to use all cells for metrics}{not only primary cells}

\tdsDoneHide{Bakker IoU comparison}{Bakker2017 uses IoU (J) as the metric, not mean IoU. Direct comparison not possible unless we calculate the cell class IoU. Bakker claims \SI{0.7}{}, \SI{0.65}{} for DISCO cell class IoU, CellSerpent and CellX around \SI{0.6}{} and Cellstar \SI{0.4}{}, the latter three have a similar run-time}

\tdsCamera{outlook pipeline after all?}{}


%

\tdsDoneHide{no:move the runtime DISCO table to discussion?}{}

%


\section{Conclusion}
\label{sec:conclusion}
\addcontentsline{tdo}{todo}{\textbf{Conclusion}}

\tdsDoneHide{change focus from dataset to segmentation of TLFM }{for sybnio etc}

\tdsDoneHide{add image cytometry emphasis}{}
In summary, we trained, validated and tested U-Net architectures for multiclass segmentation of time-lapse microscopy imagery of individual yeast cells in microstructured environments. The models discern between the geometrically similar trap microstructure and yeast cell classes, mitigating some potential sources of measurement error. S{\o}rensen-Dice $S$ of $0.96$ and mean Jaccard index $\bar{J}$ of $0.89$ is achieved. The proposed method is robust in the specific configuration it and the training dataset is designed for. Further extending the training dataset, for example by including alternate trap geometries or droplets instead of cells, promises to make the method applicable in a wide variety of quantitative \emph{in vivo} and \emph{in vitro} microstructured time-lapse fluorescence microscopy settings.




The proposed method achieves significantly improved cell segmentation performance in comparison to the reported state-of-the-art for the specific application of hydrodynamically trapped yeast imagery. We achieved an intersect-over-union for the cell class segmentation of $0.82$. This promises to reduce measurement uncertainty, facilitate cell tracking efficacy and increase the experimental data yield in future applications. We reduced the floating point operations required for a forward pass of the U-Net, while effectively maintaining performance on the yeast-trap configuration. The resulting runtimes and accurate segmentation make future online monitoring feasible, for example for closed-loop optimal experimental control. 


\tdsDoneHide{include dataset mention as result of paper}{}
\vspace{2mm}
\section*{Acknowledgements}
\label{sec:acknowledgements}
\addcontentsline{tdo}{todo}{\textbf{sec:acknowledgements}}
We thank Jan Basrawi  for contributing to data annotations and Markus Baier for aid with the computational setup.
\tdsCamera{funding acknowledgement where? title page opt available}{see IEEE thanks and very 2nd line of main}

This work was supported by the Landesoffensive f\"{u}r wissenschaftliche Exzellenz as part of the LOEWE Schwerpunkt CompuGene. H.K. acknowledges support from the European Research Council (ERC) with the  consolidator grant CONSYN (nr. 773196).
\vspace{2mm}
\tdsCamera{submission HK Q add acknowledgement of Spectra?}{}
\tdsCamera{double check Acknowledgement at end}{}
\tdsCamera{References check correct formatting for IEEE!}{}

\bibliographystyle{IEEEtran}
\bibliography{IEEEabrv,unet_bib}




\end{document}